\def\Z#1{_{\lower2pt\hbox{$\scriptstyle#1$}}}
\def\etal{{\it et al}.}   \def\DE{\Delta}
\def\dd{{\rm d}} \def\ds{\dd s}  
\def\th{\theta} \def\ph{\phi}   
\def\goesas{\mathop{\sim}\limits}  \def\gam{\gamma}
\font\sevenrm=cmr7 \def\ws#1{_{\hbox{\sevenrm #1}}} 
\def\w#1{\,\hbox{#1}} \def\kmsMpc{\w{km}\;\w{sec}^{-1}\w{Mpc}^{-1}}
 \def\SS{${\cal S}$}
 \def\tc{\tau} \def\Om{\Omega_m} \def\Ol{\Omega_{\Lambda}}
 \def\Hc{H\Z0} \def\LCDM{$\Lambda$CDM} \def\si{\sigma}
\def\sLCMD{$\scriptstyle\Lambda$CDM}
\def\Deriv#1#2#3{{#1#3\over#1#2}} \def\Oc{\widetilde\Omega\Z0}
\def\atil{\widetilde a} \def\ts{t} \def\aphys{a}
\def\beq{\begin{equation}}
\def\eeq{\end{equation}}
\def\bea{\begin{eqnarray}}
\def\eea{\end{eqnarray}}
\def\name{Fractal Bubble Model}
\def\dlw{(Wiltshire 2005a)} \def\Model{FB model}
\def\Kolb{Kolb \etal\ (2005)}
\def\placefig#1#2{\centerline{\includegraphics[width=#2truemm]{#1}}}
\ifx\pdfximage\UnDeFiNeD
\ifx\pdfimage\UnDeFiNeD
\def\deltamu{\centerline{\scalebox{0.85}{\includegraphics{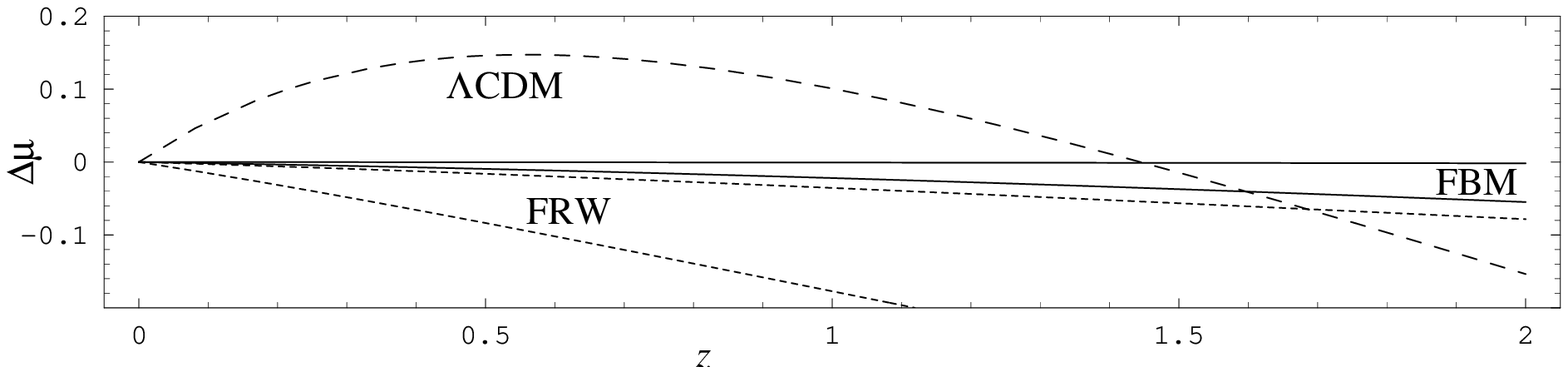}}}}
\def\lcdm{\centerline{\scalebox{0.55}{\includegraphics{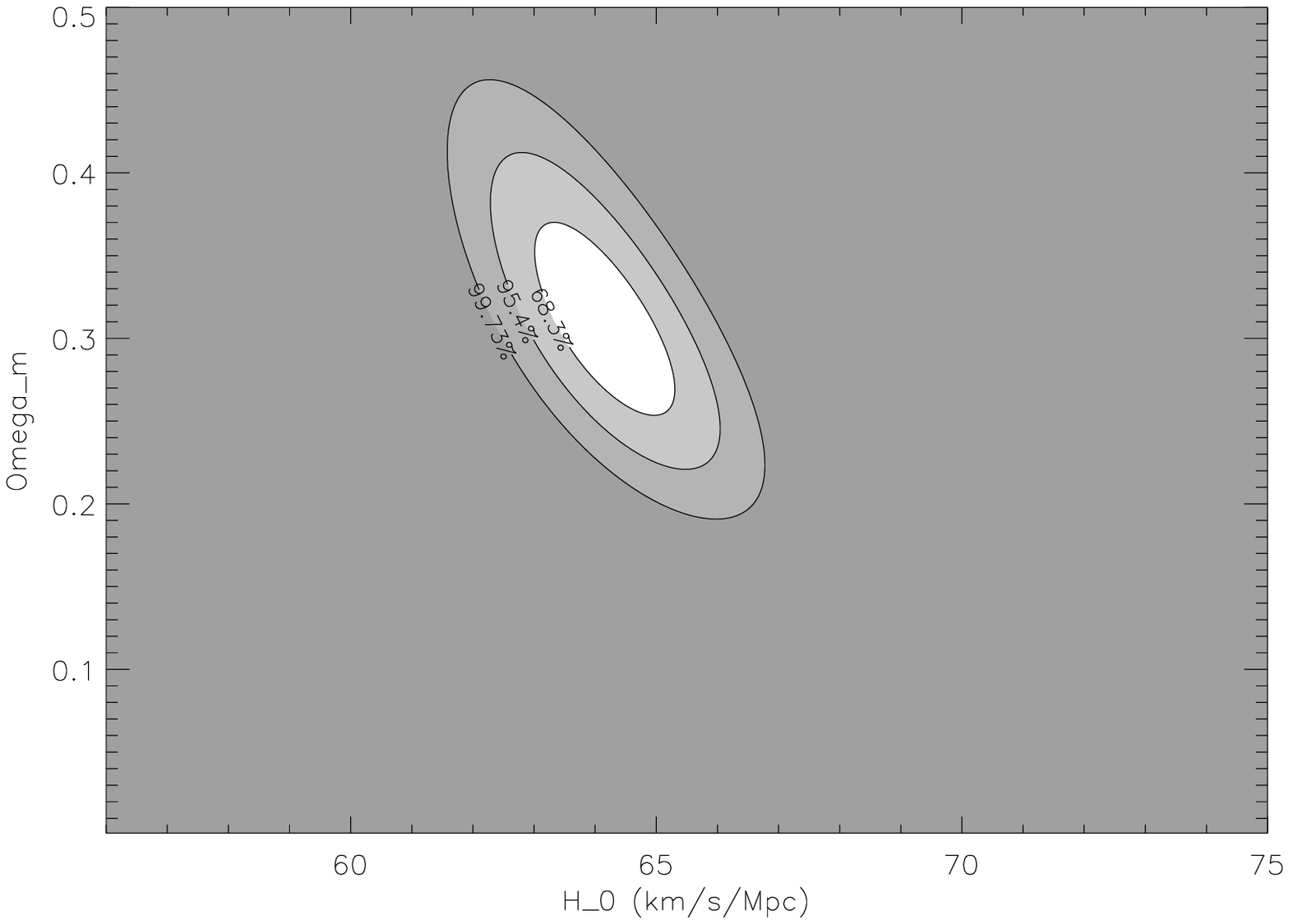}}}}
\def\obm{\centerline{\scalebox{0.55}{\includegraphics{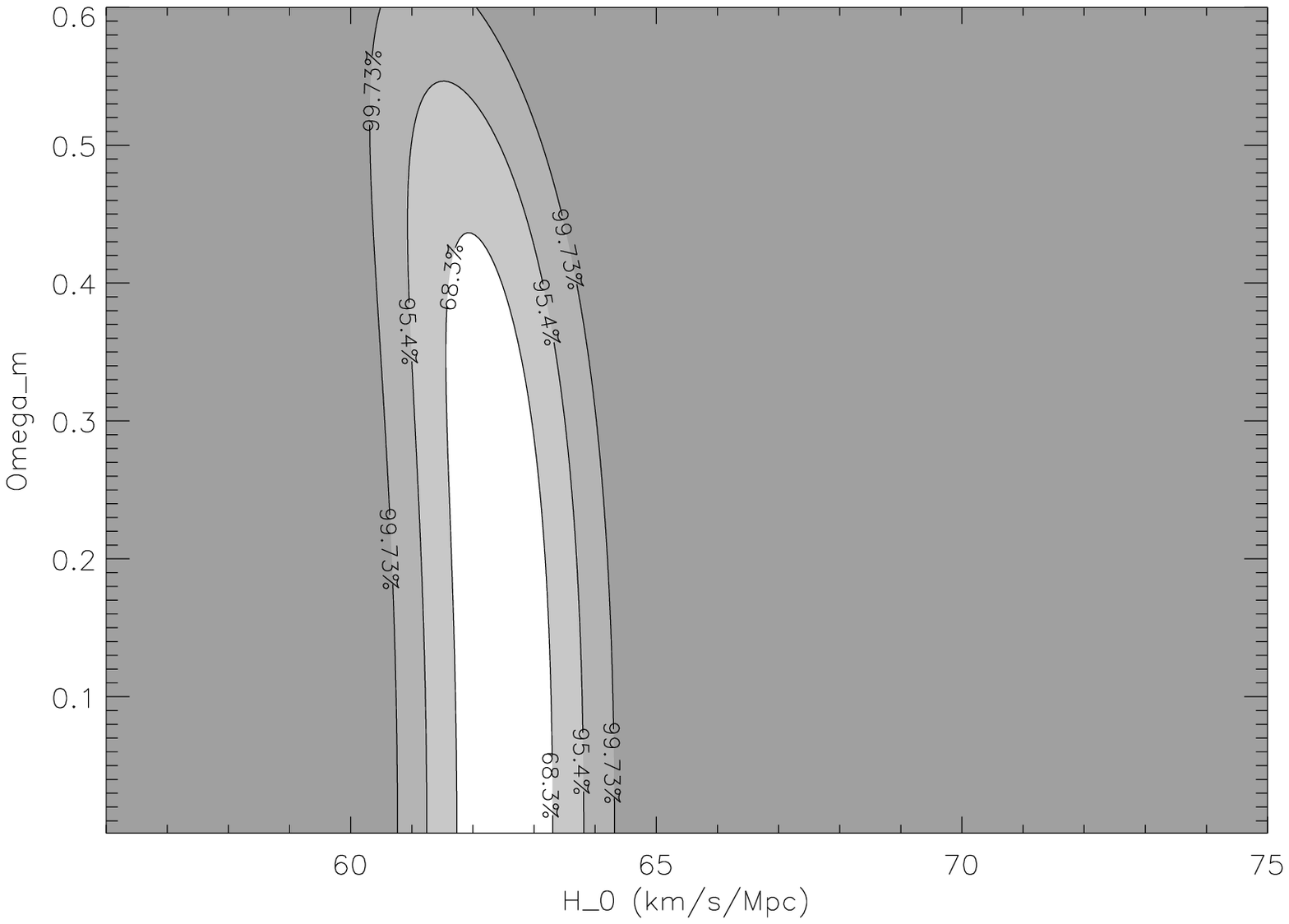}}}}
\def\slice{\centerline{\scalebox{0.85}{\includegraphics{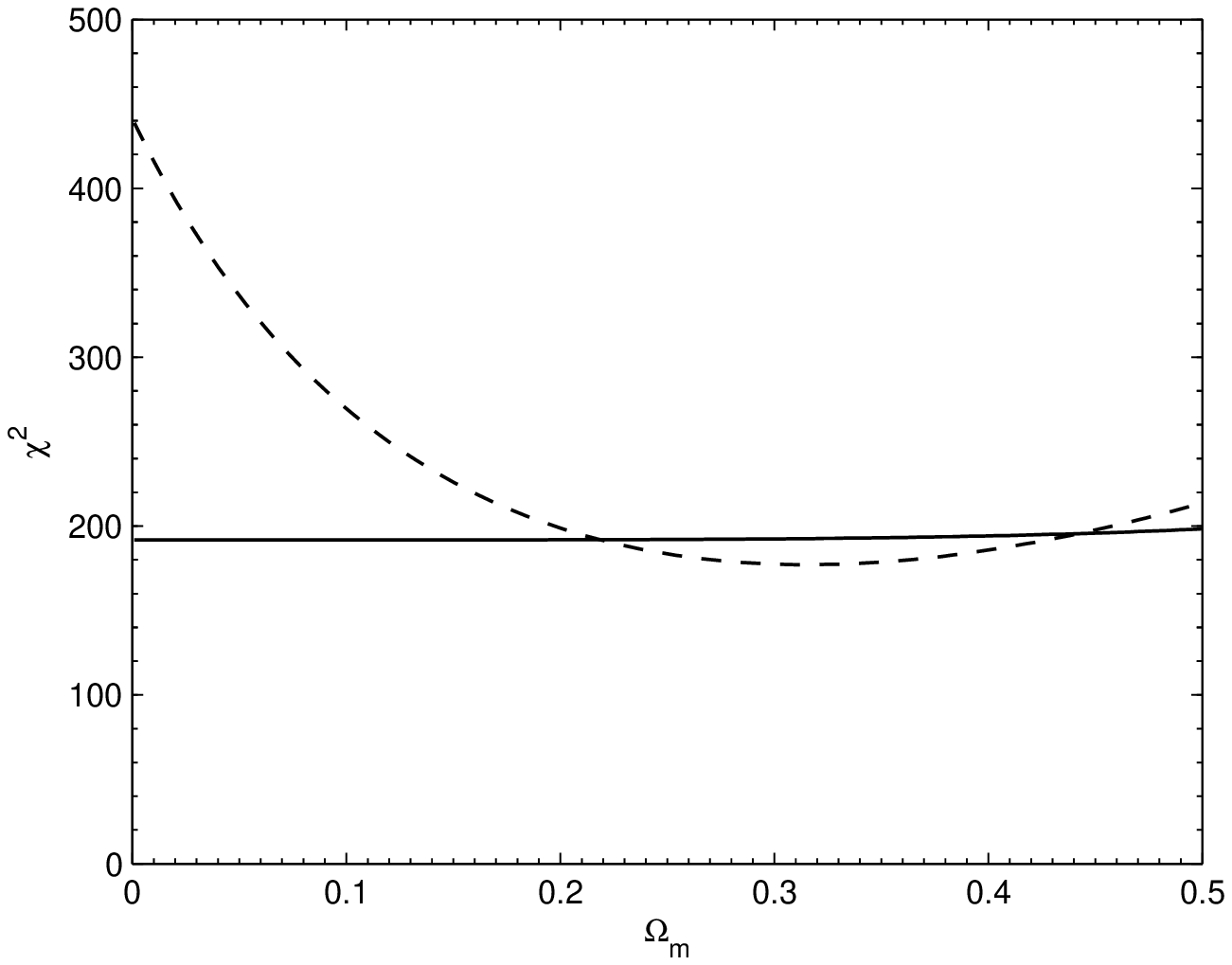}}}}
\def\bayes{\centerline{\scalebox{0.55}{\includegraphics{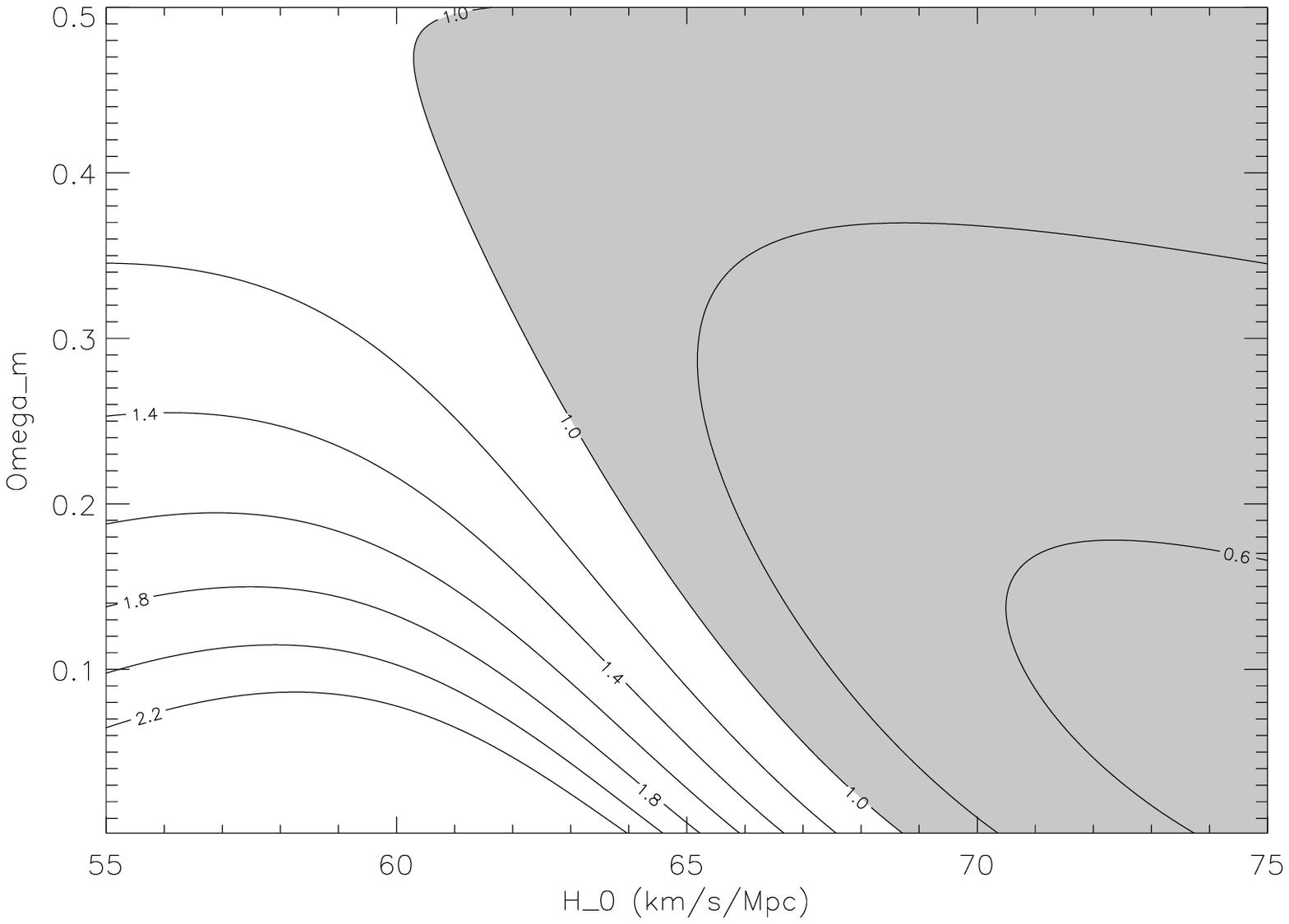}}}}
\else
\def\deltamu{\placefig{deltamu.png}{180}}
\def\lcdm{\placefig{lcdm.png}{100}}
\def\obm{\placefig{fb.png}{100}}
\def\slice{\placefig{slice.png}{100}}
\def\bayes{\placefig{bayesplot.png}{100}}
\fi
\else
\def\deltamu{\placefig{deltamu.png}{180}}
\def\lcdm{\placefig{lcdm.png}{100}}
\def\obm{\placefig{fb.png}{100}}
\def\slice{\placefig{slice.png}{100}}
\def\bayes{\placefig{bayesplot.png}{100}}
\fi
\documentclass[preprint]{aastex}%
\received{}
\accepted{}
\journalid{}{}
\articleid{}{}
\shortauthors{Carter et al.}
\shorttitle{}

\begin{document}
\title{Type Ia supernovae tests of fractal bubble universe with
no cosmic acceleration}

\author {{Benedict~M.N.~Carter\altaffilmark{1}},
{Ben~M.~Leith\altaffilmark{1}}, {S.C.~Cindy~Ng\altaffilmark{2}},
{Alex~B.~Nielsen\altaffilmark{1}}, and {David~L.~Wiltshire\altaffilmark{1}}}

\altaffiltext{1}{Department of Physics \& Astronomy,
University of Canterbury, Private Bag 4800, Christchurch, New Zealand}

\altaffiltext{2}{Physics Department, National University of Singapore,
2 Science Drive 3, Singapore 117542}

\begin{abstract}
The unexpected dimness of Type Ia supernovae at redshifts $z\la 1$ has over
the past 7 years been seen as an indication that the expansion of the
universe is accelerating. A new model cosmology, the ``fractal
bubble model'', has been proposed by one of us, based on the idea that our
observed universe resides in an underdense bubble remnant from a primordial
epoch of cosmic inflation, together with a new solution for averaging in an
inhomogeneous universe. Although there is no cosmic acceleration, it is
claimed that the luminosity distance of type Ia supernovae data will
nonetheless fit the new model, since it mimics a Milne universe at low
redshifts. In this paper the hypothesis is tested statistically against
the available type Ia supernovae data by both chi--square and Bayesian
methods. While the standard model with cosmological constant $\Ol=1-\Om$
is favoured by a Bayesian analysis with wide priors, the comparison depends
strongly on the priors chosen for the density parameter, $\Om$. The fractal
bubble model gives better agreement generally for $\Om<0.2$. It also gives
reasonably good fits for all the range, $\Om=0.01$--$0.55$, allowing the
possibility of a viable cosmology with just baryonic matter, or
alternatively with both baryonic matter and additional cold dark matter.
\end{abstract}

%\pacs{98.80.-k, 98.80.Es, 98.80.Cq, 98.80.Bp%

\section{Introduction}

For nearly a decade it has been assumed that the universe is presently
undergoing a period of acceleration driven by an exotic form of dark
energy. In a new model proposed by one of us \dlw,
it has been claimed that it is possible to fit type Ia supernovae (SNeIa)
luminosity distances without exotic forms of dark energy, and without
cosmic acceleration. The new model was developed on the basis of a
suggestion of Kolb, Matarrese, Notari and Riotto (2005), that
cosmological evolution must take into account super--horizon sized
remnant perturbations from an epoch of primordial inflation.

The new model cosmology of Wiltshire (2005a,2005b) -- henceforth the
Fractal Bubble (FB) Model --
shares the basic assumption of \Kolb\ that the observed universe resides
within an underdense bubble, \SS, in a bulk universe which has the geometry of
a spatially flat Friedmann--Robertson--Walker (FRW) geometry on the
largest of scales. In other respects, however, the \Model\
differs substantially from that of \Kolb. In particular,
the \Model\ is based on exact solutions of Einstein's
equations, whereas \Kolb\ based their analysis on linearized
perturbations, with a number of approximations. Furthermore, while
Kolb \etal\ suggested that the deceleration parameter would behave as
$q(\tc)\to-1$ at late times, in the \Model\
$q(\tc)\to0$. The \Model\ has the added advantage that
it can make quantitative predictions for many cosmological quantities
in the epoch of matter domination
based on two parameters, the Hubble constant, $\Hc$, and the density
parameter, $\Om$.

The \Model\ makes two crucial physical assumptions. The first,
as just mentioned, is that our observed universe resides in an
underdense bubble, which is a natural outcome of primordial inflation.
The second assumption is that on account of the bottom--up manner in which
structure forms and spatial geometry evolves in an inhomogeneous universe
with the particular density perturbations that arise from primordial
inflation, the local clocks of isotropic observers in average galaxies in
bubble walls at the present epoch, are defined in a
particular way. The small scale bound systems which correspond to typical
stars in typical galaxies retain local geometries with an asymptotic time
parameter frozen in to match
the cosmic time of the average surfaces of homogeneity in their past light
cones at the epoch they broke away from the Hubble flow. That local
average geometry was a spatially flat Einstein--de Sitter
universe, even though the presently observable universe, \SS, is
underdense when averaged on the larger spatial scales where structure forms
later. These arguments are clarified at length by Wiltshire (2005b).

The physical implication of this is that we must consider an
alternative solution to the fitting problem in cosmology
(Ellis and Stoeger 1987). The open FRW geometry
\beq\label{FRWopen1}
\dd\widetilde s^2 = - \dd\ts^2 + \atil^2(\ts)
\left[{\dd r^2\over1+r^2}+r^2(\dd\th^2+\sin^2\th\,\dd\ph^2)\right],
\eeq
retains meaning as an average geometry for the whole universe at late
epochs, but most closely approximates the local geometry only in voids, these
being the average spatial positions on spatial hypersurfaces. However, in
the matching of asymptotic geometries between those scales and the
non--expanding scales of bounds stars and star clusters in
average galaxies in bubble walls, it is assumed that a non-trivial
lapse function $\gam(\tc)=\Deriv{\dd}\tc{\ts}$ enters, so that observers
in such galaxies describe the same average geometry (\ref{FRWopen1}) in
terms of a different conformal frame
\beq\label{FRWopen3}
\dd\widetilde s^2 = \gam^2(\tc)\ds^2\,,
\eeq
where
\beq\label{FRWopen4}
\ds^2= -\dd\tc^2 + \aphys^2(\tc)
\left[{\dd r^2\over1+r^2}+r^2(\dd\th^2+\sin^2\th\,\dd\ph^2)\right],
\eeq
and $\aphys\equiv \gam^{-1}\atil$. The lapse function is uniquely determined,
and with a careful re--calibration of cosmic clocks,
a model cosmology can be constructed, which has the potential to fit
luminosity distances without cosmic acceleration.

It is the aim of this {\it Letter} to test this claim against the available
SNeIa data. We adopt the notation and conventions of Wiltshire (2005a,2005b). 

\section{Observable quantities}

Starting from the standard definition of the redshift $z$,
\beq 1+z \equiv \frac{\lambda\ws{observed}}{\lambda\ws{emitted}}\,,\eeq
the luminosity distance for the cosmological model given
in (Wiltshire 2005a)\ is
\beq d\Z{L}
=\frac{c(1+z)(2+\Oc^2)}{\Hc\Oc (2+\Oc)}
\left[2\cosh\eta-\frac{2-\Oc}{\sqrt{1-\Oc}}\sinh\eta\right]\,,
\label{dl}
\eeq
where $\Hc$ is the currently measured value of the Hubble constant,
$\Oc$ is a conveniently defined parameter, related to the current
matter density parameter, $\Om$, according to
\beq
\Oc={6\over\sqrt{\Om}}\sin\left[{\pi\over6}-{1\over3}\cos^{-1}\sqrt{\Om}
\right]-2\,,
\eeq
and $\eta$ is given by
\beq
\cosh\eta = -\frac{1}{2} + \frac{(1-\Oc)(2+\Oc)
+ \sqrt{\Oc z[9\Oc z - 2\Oc ^{2}+16\Oc +
4] + (\Oc ^{2}+2)^{2}}}{2\Oc (z+1)}\,.
\eeq
%

%------------------------------------------------------------
\begin{figure}[h]
\vbox{\deltamu
\caption{\label{fig:deltamu} {\sl Difference of distance modulus from
that of an empty (Milne) universe, $\DE\mu$, versus redshift. The dashed line
corresponds to the \LCDM\ model with $\Om=0.27$, $\Ol=0.73$; the solid lines
correspond to the \name\ (FBM); and the dotted lines corresponds
the open, matter only, universe (FRW). For the FBM and FRW models, in each
case the upper line corresponds to $\Om=0.05$, the lower line to $\Om=0.27$.
($\Hc=62.7\kmsMpc$ in all cases.) For $\Om=0.05$ FB line is indistinguishable
from the horizontal axis over the range shown.}}}
\end{figure}
%------------------------------------------------------------

\section{Data analysis}

To make contact with observation the luminosity distance (\ref{dl})
is related to the distance modulus $\mu$ by the standard
formula
\beq \mu \equiv m-M = 5\log_{10}(d\Z{L}) + 25\,.\eeq
Let us define $\DE\mu = m\ws{model} - m\ws{empty}$, where $m\ws{empty}$ is
the apparent magnitude for %an open and totally empty
the $\Om=0$ universe. A plot of $\DE \mu$ versus $z$ for the
\Model\ as compared to the \LCDM\ and open FRW models is given in
Fig.~\ref{fig:deltamu}. For $\Om=0.05$ the distance modulus of the FB
model is indistinguishable from that of a Milne universe over the range
$0\le z\le2$; while even the $\Om=0.27$ FB model is closer to a Milne
universe than the $\Om=0.05$ open FRW model.

\begin{figure}[t]
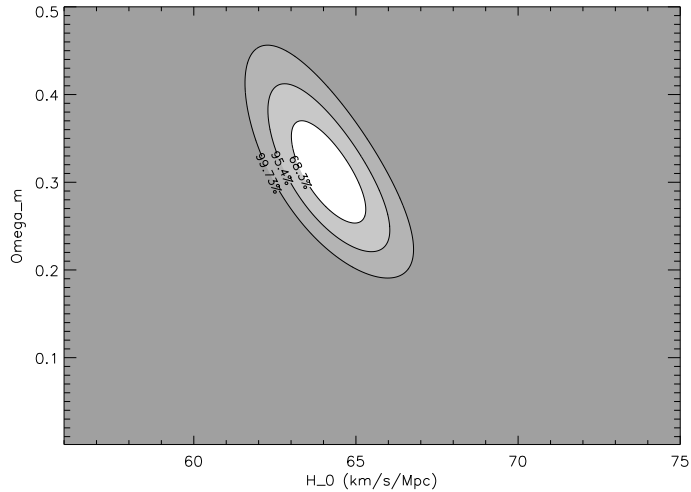

\vbox{\lcdm
\caption{\label{fig:lcdm}%
{\sl Confidence limits for the estimation of parameters for the
$\Lambda$CDM model, with $\Ol=1-\Om$.}}}
\end{figure}

\begin{figure}[h b]
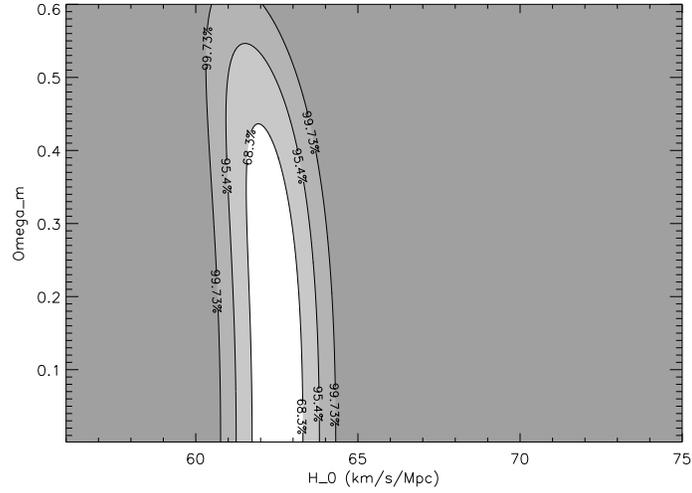

\vbox{\obm
\caption{\label{fig:fb}%
{\sl Confidence limits for the estimation of parameters for the \name\ .}}}
\end{figure}

We have compared the new \Model\ to the supernova data using
the ``Gold data set'' recently published by Riess \etal\ (2004).
The $\chi^{2}$ and Bayes factor method employed followed the one outlined
in Ng and Wiltshire (2001). Use of the luminosity distance (\ref{dl})
requires a series expansion near $\Oc=0$ to avoid numerical problems:
for $\Oc<0.02$ we used a Laurent series to order $9$. The results
are presented here in Figs.~\ref{fig:lcdm} and \ref{fig:fb}, where
the 1$\si$, 2$\si$ and 3$\si$ confidence contours are plotted
in the $(\Om,\Hc)$ parameter space.

%------------------------------------------------------------
%------------------------------------------------------------
%------------------------------------------------------------
%------------------------------------------------------------
%------------------------------------------------------------
\begin{figure}[t]
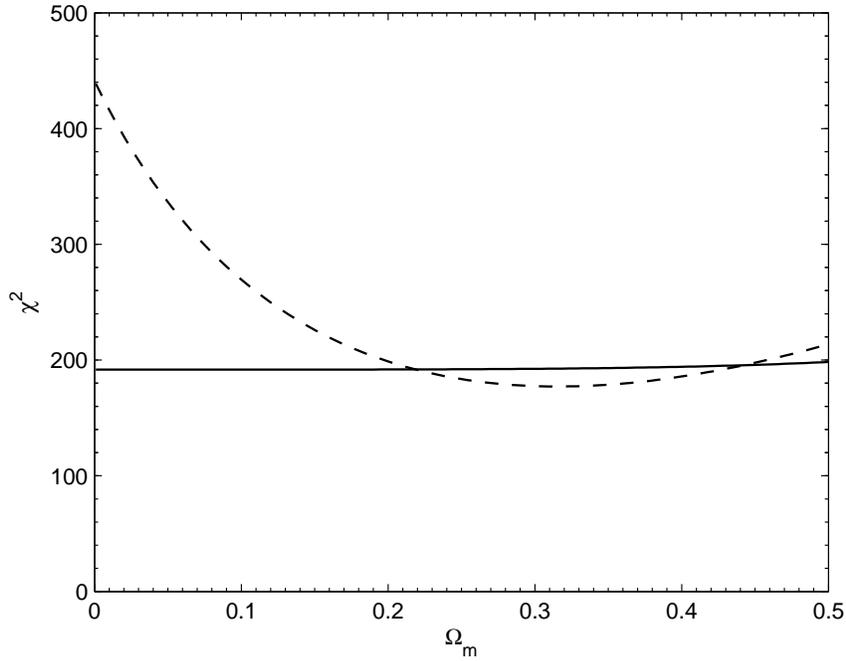

\vbox{\slice
\caption{\label{fig:slice}%
{\sl A slice of the confidence plot for the \name\ at its best--fit
value $\Hc=62.683$ (solid line) for $0\leq\Om\leq0.5$, and for the \LCDM\
model at its best--fit value $\Hc=64.145$ (dashed line).}}}
\end{figure}
%------------------------------------------------------------

A good fit for the $\Lambda$CDM model was at $\Hc= 65\kmsMpc$ and
$\Om = 0.27$ with ($\chi^{2}=178$) (Riess \etal\ 2004). For the
\name\ the best fit values are at $\Hc= 62.7^{+0.6\,(1\si),\;+1.1\,(2\si)}
_{-1.1\,(1\si),\;-1.7\,(2\si)}\kmsMpc$ and include the $1\si$ range
$\Om<0.44$ ($\chi^{2}=191.7$). The value of $\chi^2$ is affected so
little by the value of $\Om$ in this range that we do not quote a
single best--fit value. The $2\si$ confidence limit is $\Om<0.55$.
$\Om$ can be made arbitrarily small within the $1\si$ bounds.
Since recalibrated primordial nucleosynthesis
bounds for baryonic matter are typically in the range
$\Omega_b\goesas0.02$--$0.13$, we see that the supernova data fit well for
the \Model\ regardless of whether or not there is cold dark matter in
addition to baryonic matter.

A Bayes factor for the two models can be calculated from the data by
\beq B_{12} = \frac{\int\dd \Hc\,\int\dd\Om \exp(-\chi\ws{model
1}^{2}/2)}{\int\dd \Hc\,\int\dd\Om \exp(-\chi\ws{model 2}^{2}/2)}\,.
\label{bayes}\eeq
This result can be interpreted using Table \ref{tab:bayesfac}. For
very wide priors, namely $0.001\le\Om\le0.5$ and $55\le \Hc\le75\kmsMpc$,
the Bayes factor is 396 in favour of \LCDM. However, we 
note that the Bayes factor is very sensitive to
the range chosen for the priors, since the best--fit values for the two
models are include quite distinct values of $\Om$. This is seen in Table
\ref{tab:bayescomp}, where narrower priors are used for $\Hc$, and a
variety of priors for $\Om$. The examples with baryonic matter only
use values of the baryon matter density from primordial nucleosynthesis
bounds, as specifically recalibrated for the FB model by Wiltshire (2005b):
it is found that the density fraction of ordinary baryonic
matter is 2--3 times that which is conventionally obtained, and typically
$\Omega_b\goesas0.1$ is expected.
%------------------------------------------------------------
\begin{table}[hb]
\centering
\caption{Interpretation of Bayes Factors}
\bigskip
\begin{tabular}{l|l}
$B_{12}$ & Strength of evidence for $H_1$ over $H_2$ \\
\hline
1 to 3 & Not worth more than a bare mention\\
3 to 20 & Positive\\
20 to 150 & Strong\\
$>150$ & Very Strong\\
\label{tab:bayesfac}
\end{tabular}
\end{table}
%------------------------------------------------------------
%------------------------------------------------------------
\begin{table}[hb]
\centering
\caption{Bayes Factor comparison of a $\Ol=1-\Om$ \LCDM\ model versus
the \Model\ with a fixed prior for $\Hc=100h\kmsMpc$,
$58\le h\le72$, and varying priors for $\Om$. The baryonic density
parameter bounds for the \Model\ are determined by Wiltshire (2005b).}
\bigskip
\begin{tabular}{l|l|l|l}
$\Om$ prior& Model type& $B\ws{(\sLCMD):(FBM)}$ & Model favoured\\
\hline
$0.01\le\Om\le0.5$&Wide priors&396&\LCDM\\
$0.2\le\Om\le0.3$&favoured CDM range&649&\LCDM\\
%$0.15\le\Om\le0.35$&wider CDM range&722&\LCDM\\
$0.3\le\Om\le0.5$&high density CDM&1014&\LCDM\\
$0.01\le\Om\le0.2$&low density CDM&0.53&FBM (slightly)\\
$0.08\le\Om\le0.13$&baryonic matter, low D/H&$2.6\times10^{-5}$&FBM\\
$0.02\le\Om\le0.06$&baryonic matter, high D/H&$1.3\times10^{-14}$&FBM\\
\label{tab:bayescomp}
\end{tabular}
\end{table}
%------------------------------------------------------------
\begin{figure}[t]
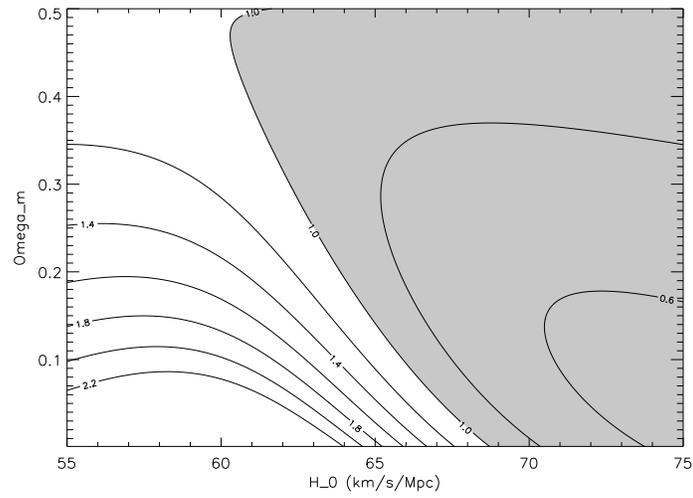

\vbox{\bayes
\caption{\label{fig:bayesplot}%
{\sl A contour plot of $\chi^2\ws{FBM}/\chi^2\ws{\sLCMD}$ 
for \name\ vs the $\Lambda$CDM model. The \name\ 
fares better in the lower left region (white), while the
\LCDM\ model is favoured in the upper right region (lightly shaded).}}}
\end{figure}
%------------------------------------------------------------

To characterise the parameter space better,
we also give a plot of the $\chi^2$ value as a function of $\Om$ for the
best--fit value of $\Hc$ in each model in Fig.~\ref{fig:slice}, and
similarly plot $\chi^2\ws{FBM}/\chi^2\ws{\sLCMD}$ for the
two models for the whole parameter space spanned by $\Om$ and $\Hc$
in Fig.~\ref{fig:bayesplot}. The \name\ is favoured for
lower values of $\Om$ and $\Hc$, and \LCDM\ for higher values of both
parameters. Also observe from Fig.~\ref{fig:slice} how insensitive the
\Model\ is to the value of $\Om$ in the range $0<\Om\la0.5$.
\section{Data reduction}

While SNeIa serve as excellent standard candles there is some
dispersion observed in their absolute magnitudes. However, a
relation first proposed by Phillips (1993), relates the
absolute magnitude, $M$, to the change in apparent magnitude from
peak to 15 days after peak, $\DE m_{15}$, as measured by clocks
{\it{in the rest frame of the supernova}}. Various methods
have been used to fit light curves to obtain absolute magnitudes
(Drell, Loredo \& Wasserman 2000, Tonry 2003).

It is also well demonstrated that redshift causes a broadening of light
curves (Leibundgut 1996) corresponding to the phenomenon of cosmic
time dilation. In standard cosmology the redshift is entirely
attributable to cosmic expansion. It is then a simple matter to
transform the observed light curve to the rest frame. In the
\Model
\beq 1+z =
\frac{\atil\Z0\gam(\tc\ws{emit})}{\atil(\tc\ws{emit})\gam\Z0}\,, \eeq
so that the measured redshift differs from that inferred from (\ref{FRWopen1}).
However, for null geodesics the transformation to the rest frame only
involves only the overall $(1+z)$ factor of the measured redshift, so that
the light curve broadening determination will be unaffected.
Only measurements which involved some independent determination of
$\gam(\tc\ws{emit})/\gam\Z0$, would lead to measurable differences between
this model and models based on the conventional identification of clock
rates.

It is impossible to comment on the Phillips relations, and other related
aspects of data--fitting, without access to the raw data. However, since these
are relations are empirical, it is not likely that there would be significant
differences in the \Model.

The only likely systematic errors that are as yet unaccounted for in
reducing the data for comparison with the \Model\ are those effects
directly related to the inhomogeneous geometry. The metric (\ref{FRWopen1})
represents an average geometry only, and for the redshifts which presently
constitute the bulk of the sample, a better understanding of intermediate
scales in the fitting problem may be required. Studies of local voids --
at scales well within the bubble \SS -- do reveal that measurable effects
on cosmological parameters are possible (Tomita 2001). In the present
model, the average rate of expansion of typical $30h^{-1}$ Mpc voids would
differ from that of much larger voids. While the dominant effect on
luminosity distances suggested here is a new effect due to the
identification of an alternative homogeneous time parameter in solving the
``fitting problem'', further scale dependent corrections in the
fractal geometry are certainly to be expected.

\section{Conclusion}
From the standard Bayesian analysis with wide priors we would conclude
that on the basis of the ``Gold data set'' of (Riess \etal\ 2004), the
standard \LCDM\ model is very strongly favoured over the \Model.
Nonetheless, this conclusion depends strongly on the priors
assumed, as illustrated by Table \ref{tab:bayescomp} and Figs.~\ref{fig:slice}
and \ref{fig:bayesplot}. We must also bear in mind that the average geometry
(\ref{FRWopen1})--(\ref{FRWopen4}) is just a first approximation to the
fitting problem. A better understanding of the differential expansion rates
of different characteristic sizes of voids in
the inhomogeneous geometry may improve the fit of the \Model\ at those
redshifts which are most significant for the claims of cosmic acceleration,
rather than a nearly empty universe deceleration. The \Model\
has other significant advantages in terms of it greatly increased
expansion age (Wiltshire 2005a), which has the potential to explain
the formation of structure at large redshifts. Thus it should be
considered as a serious alternative to models with dark energy,
quite apart from its intrinsic appeal of being a model based
simply on general relativity and primordial inflation without needing
exotic dark energy, the presence of which at the present epoch would be
a mystery to physics.

In the end, nature is the final arbiter, and future supernova observations
at higher redshifts, such as those of the SNAP mission, will certainly be
able to much more decisively distinguish between the \Model\ and \LCDM,
given the significantly more rapid deceleration of \LCDM\ at redshifts
$z>2$.

\medskip\noindent
{\bf Acknowledgements}: This work was supported by the Marsden Fund of
the Royal Society of New Zealand. We thank Michael Albrow for helpful
comments.

\medskip\noindent
{\bf Note added}: After this work appeared an independent SneIa data
analysis of the Milne universe -- considered as an approximation to a variety
of alternative theories -- has been performed (Sethi, Dev \& Jain 2005).
They obtain a comfortable fit just as we would expect from our results
given that the FB model mimics a Milne universe for $z<2$.

\end{document}